\documentclass[preprint,tightenlines,aps]{revtex4}
\begin{document}
\title{$B$ meson spectroscopy}
\author{J. Vijande$^{1,2}$,  A. Valcarce$^1$, F. Fern\'{a}ndez$^1$}
\affiliation{$^1$ Departamento de F\'{i}sica Fundamental and IUFFyM,
Universidad de Salamanca, E-37008 Salamanca, Spain.\\
$^2$ Departamento de F\' \i sica Te\'orica - IFIC,
Universidad de Valencia - CSIC, E-46100 Burjassot, Valencia, Spain.}
\vspace*{1cm} 
\begin{abstract}
We study the $B$ meson spectroscopy allowing the 
mixture of conventional $P$ wave 
quark-antiquark states and four-quark components.
A similar picture was used to describe 
the new $D_J$ and $D_{sJ}$ open charm mesons.
The four-quark components shift the masses of some
positive parity $B_{sJ}$ states
below their corresponding isospin preserving 
two-meson threshold and therefore they are expected to 
be narrow. Electromagnetic decay widths are analyzed. 

\vspace*{2cm} 
\noindent Keywords: Bottom mesons, Bottom-strange mesons, quark models.
\newline
\noindent Pacs: 14.40.Lb, 14.40.Ev, 12.39.Pn. 
\end{abstract}
\maketitle
\newpage

Heavy-light mesons play in QCD a similar role as the hydrogen 
atom in QED. This analogy provides a simple way to make predictions 
for their excited states. 
In the limit $M_Q\rightarrow\infty$ heavy-light mesons can be 
characterized by the spin of the heavy quark, $S_Q$, the total angular momentum of 
the light quark, $\vec{j_q}=\vec{S_q}+\vec{L}$, and the total angular momentum, 
$\vec{J}=\vec{S_Q}+\vec{j_q}$. 
For $P$ wave excited states there appear two degenerate doublets:
one corresponding to $j_q=1/2$, and the other to $j_q=3/2$, with quantum numbers 
$J^P=0^+, 1^+$ and $J^P=1^+, 2^+$, respectively. Those 
states with $j_q=1/2$ can only decay via an $S$ wave transition, whereas the $j_q=3/2$ states 
undergo a $D$ wave transition. Therefore the decay widths are expected to be much 
broader for $j_q=1/2$ than $j_q=3/2$ states. We denote the $J^P=(0^+,1^+)_{j_q=1/2}$ 
states as $(A^*_0,A^{'}_1)$ and the $J^P=(1^+,2^+)_{j_q=3/2}$ states as $(A_1,A^*_2)$,
with $A=B$ (for $b\overline{n}$ states, where $n$ stands for a light $u$ or $d$ quark) 
or $B_s$ (for $b\overline{s}$ states). 
Based on this description the expected decay properties of the 
$P$ wave $B_{sJ}$ mesons are rather simple, they are summarized in 
Table~\ref{t1}.

A few predictions of masses and widths are available from theory 
\cite{Ebert,God,Bardeen,col,Falk,Eich} and
lattice simulations~\cite{green}, see Table~\ref{t2}. Similar 
masses and widths of the $L=1$ excited state are predicted, 
around $5.7$ GeV for $B$ 
and $5.8$ GeV for $B_s$, except for Refs.~\cite{Bardeen}
and~\cite{col} reporting lower masses for the $B^*_{s0}$ and $B^{'}_{s1}$. These 
works use model parameters fitted on 
the $D_s$ meson sector and, therefore, they incorporate the anomalies 
we will comment below.

From the experimental point of view the spectroscopy of excited mesons 
containing bottom quarks is still not well known. Only the ground $0^-$ and 
the excited $1^-$ states are well established in the PDG~\cite{PDG}.
There are some experimental data by the L3 
Collaboration \cite{L3} reporting the first measurement 
of the $B'_1$ and $B^*_2$ masses, 5670$\pm10\pm13$ MeV
and 5768$\pm5\pm6$ MeV, respectively. Recently,
D0 and CDF Collaborations have reported results on the 
spectroscopy of orbitally excited bottom mesons \cite{Pau}. CDF found two 
states, $B_1$ and $B^*_2$, with masses $M(B_1)=5734\pm3\pm2$ MeV and 
$M(B^*_2)=5738\pm6\pm1$ MeV.
D0 also found the same states but with slightly different masses,
$M(B_1)=5720.8\pm2.5\pm5.3$ MeV and $M(B^*_2)-M(B_1)=25.2\pm3.0\pm1.1$ MeV. 
In the strange sector CDF reported two narrow $B_{s1}$ and $B^*_{s2}$ states
with masses $M(B_{s1})=5829.4$ MeV and $M(B^*_{s2})=5839$ MeV while D0 measured
only the $B^*_{s2}$, with a mass of $5839.1\pm1.4\pm1.5$ MeV.

The poorly known experimental situation in the open-beauty sector is far from
the one observed for open-charm mesons. 
Since 2003, when the $D^*_{sJ}(2317)$ and the $D_{sJ}(2460)$
were discovered by BABAR Collaboration~\cite{BaBar}, eight $c\bar s$ new states have 
been reported, more than needed to fill the $L=0$ 
doublet and the four $L=1$ states. Also the number of $c\bar n$ states has 
grown with the Belle 
observation \cite{Belb4} of a broad scalar resonance, $D^*_0$,
with a mass of $2308\pm 36$ MeV
and a width $\Gamma=276 \pm 66$ MeV.
Some of these new states present unexpected masses, quite 
different from those predicted by quark potential models if 
a pure $c\overline{q}$ configuration is considered. If they would 
correspond to standard $P$ wave mesons, their masses would be similar 
to the already known $L=1$ $c\overline{q}$ states, namely 
around 2.4 GeV  for $c\overline{n}$ and 2.5 GeV for $c\overline{s}$.
They would be therefore above the $D\pi$, $D^*\pi$ and $DK$, $D^*K$ 
thresholds, respectively, being broad resonances. However, 
the $D_{sJ}^*(2317)$ and $D_{sJ}(2460)$ states are below 
the $DK$ and $D^*K$ thresholds and therefore they can only decay through 
the isospin forbidden channels $D_s \pi^0$ and $D^*_s \pi^0$ with a 
very narrow width. In the case of the $D^*_0(2308)$ the large width 
observed would be theoretically expected although not its low mass.

These unexpected properties of the $D_{sJ}^*(2317)$ and the $D_{sJ}(2460)$ mesons
have been explained in Ref.~\cite{Vij73} assuming a mixture of $q\bar q$ $(L=1, S=1,0)$
and $qq\bar q\bar q$ $(L=0, S=1,0)$ components. The reason why four-quark 
configurations, which mimic quark loop contributions, are important 
in this case is that whereas the $q\bar q$ pair is in 
a $L=1$ state, the four-quark state is an $S$ wave and thus its contribution may 
be relevant.
It is well known that the 
existence of four-quark configurations is favored in the 
heavy-quark sector because, due to the coulombic character of the systems, 
the binding energy augments proportionally to the mass whereas the 
kinetic energy contribution gets reduced when the mass increases. Therefore, 
if sizeable contributions of four-quark structures appear in 
the $D$ meson sector, they should also be present for $B$ mesons.
Therefore, one may wonder which are the consequences of the mixing 
between two- and four-quark components in the $B$ meson spectra. 

In the present work we have extended the analysis done on
Ref.~\cite{Vij73} for open-charm mesons to excited $P$ wave open-beauty 
mesons. Let us first briefly resume the basic
features of the constituent quark model used \cite{Vij31}. The model includes a 
dynamical quark mass appearing as a consequence of the spontaneous
breaking of the original QCD $SU(3)_{L}\otimes SU(3)_{R}$ chiral symmetry at
some momentum scale. Once the dynamical quark 
mass is generated, whatever the mechanism, such quarks have
inevitably to interact through Goldstone modes 
such that the rotation of the quark fields can be 
compensated by the boson fields. A simple lagrangian 
invariant under the chiral transformation has been derived in Ref.~\cite{Diak96},

\begin{equation}
L=\overline{\psi }(i\partial -M(q^2)U^{\gamma _{5}})\psi\, ,
\end{equation}
where $U^{\gamma _{5}}=\exp (i\pi ^{a}\lambda ^{a}\gamma _{5}/f_{\pi })$,
$\pi ^{a}$ denotes nine pseudoscalar fields $(\eta _{0,}\vec{\pi }
,K_{i},\eta _{8})$ with $i=$1,...,4 and $M(q^2)$ is the constituent mass.
The constituent quark mass can be explicitly obtained from the theory and
parametrized as $M(q^{2})=m_{q}F(q^{2})$ with

\begin{equation}
F(q^{2})=\left[ \frac{{\Lambda}^{2}}{\Lambda ^{2}+q^{2}}
\right] ^{\frac{1}{2}} \, .
\end{equation} 
Once this has been done $U^{\gamma _{5}}$ can be expanded in terms of boson fields,

\begin{equation}
U^{\gamma _{5}}=1+\frac{i}{f_{\pi }}\gamma ^{5}\lambda ^{a}\pi ^{a}-\frac{1}{%
2f_{\pi }^{2}}\pi ^{a}\pi ^{a}+...
\end{equation}
The first term of the expansion generates the constituent quark mass while the
second gives rise to a one-boson exchange interaction between quarks. The
main contribution of the third term comes from the two-pion exchange which
has been simulated by a scalar exchange potential.

For higher momenta we assume that quarks still interact through gluon
exchanges. Following  Ref.~\cite{Ruj75} the gluon exchange can be described
as an effective interaction between constituent quarks given by 

\begin{equation}
L_{gqq}=i\sqrt{4\pi }\alpha _{s}\overline{\psi }\gamma _{\mu }G^{\mu
}\lambda _{c}\psi \, .
\end{equation}

Finally, lattice calculations in the quenched
approximation derived, for heavy quarks, a confining interaction linearly
dependent on the interquark distance. The consideration of sea quarks apart
from valence quarks (unquenched approximation) suggests a screening effect on
the potential when increasing the interquark distance~\cite{Bal01}. In terms of 
a potential, the color screening produces a 
gradual decrease of the potential slope, i.e.,

\begin{equation}
V_{CON}(\vec{r}_{ij})=\{-a_{c}\,(1-e^{-\mu_c\,r_{ij}})+ \Delta\}(\vec{%
\lambda^c}_{i}\cdot \vec{ \lambda^c}_{j})\,
\end{equation}

\noindent where $\Delta $ is a global constant to fit the origin of
energies. At short distances this potential presents a linear behavior
while it becomes constant at large distances.

Using this model we have solved the Schr\"odinger equation for the two-
and four-body problems. The two-body case has been solved exactly, while
to solve the four-body 
case we have used a variational method 
with a radial trial wave function taken as the 
most general combination of generalized gaussians~\cite{Suz98}. 
This wave function includes all possible flavor-spin-color 
channels contributing to a given configuration~\cite{tetra}. 

To describe the mixing between $q\bar q$ and four-quark components
we use the simplest version of the coupled channel model. Namely, it 
is assumed that a meson state is given by 
\begin{equation}
\label{mes-w}
\left|\psi\right>=\sum_{i} \alpha_i \left|q
\bar q\right>_i + \sum_{j} \beta_j \left|qq\bar q \bar q\right>_j
\end{equation}
where $q$ stands for quark degrees of 
freedom and the coefficients $\alpha_i$ and $\beta_j$ take into 
account the mixing. The meson systems could then be 
described in terms of a hamiltonian $H=H_0+H_1$, being
\begin{equation}
H_0 = \left( \matrix{H_{q\bar q} & 0 \cr
0 & H_{qq\bar q\bar q} \cr } \right) \,\,\,\, {\rm and}\,\,\,\,
H_1 = \left( \matrix{0 & V_{q\bar q \leftrightarrow qq\bar q\bar q} \cr 
V_{q\bar q \leftrightarrow qq\bar q\bar q} & 0 \cr } \right)\,,
\label{eq1}
\end{equation}
where $H_0$ is the constituent quark model hamiltonian described above and
$H_1$ takes into account the mixing between
$q\overline{q}$ and $qq\bar{q}\bar{q}$ configurations. It includes the
annihilation operator of a quark-antiquark pair into the vacuum, that can
be obtained from the $^{3}P_{0}$ model. Since this 
model depends on a vertex parameter, we determine this parameter
by looking to the quark pair that is annihilated (and 
not to the spectator quarks that will form the final 
$q\overline{q}$ state). Therefore we have taken 
$V_{q\overline{q} \leftrightarrow qq\bar{q}\bar{q}}=\gamma $.
If this coupling is weak enough one can solve independently 
the eigenproblem for the hamiltonians $H_{q\overline{q}}$ 
and $H_{qq\bar{q}\bar{q}}$, treating $H_{1}$ perturbatively. 
To ensure that the perturbative treatment is justified, 
$\gamma$ cannot take all possible values, being restricted to
$|\gamma/(E^{\,n}_{J^P}-E^{\,n+1}_{J^P})|^2\leq1$. This restriction
will limit the energy range of the mixed states once
the unmixed energies are calculated. 
The parameter $\gamma$ has been fixed in Ref.~\cite{Vij73} to reproduce 
the mass of the $D_{sJ}^*(2317)$, being $\gamma$=240 MeV.

Let us first compare the experimental results with the predictions of 
our model for the $b\bar n$ and $b\bar s$ quark pairs.
In Table \ref{t3} it can be seen that the model nicely reproduces 
the known experimental data.
The two states which have not been measured, ($A^*_0,A^{'}_1$) lie above 
the $BK=5774$ MeV and $B^*K=5820$ MeV thresholds for the strange 
sector, and above the $B\pi=5417$ MeV threshold for the non-strange one. 

Once the mixing of the $q\bar q$ pairs with the $bq\bar q\bar q$ states is considered,
the $J^P=0^+$ and $1^+$ states acquire   
almost a 30\% of four-quark component (see Table~\ref{t4}).
Without being dominant, this component
is the responsible for shifting the mass of the 
unmixed states below the $BK$ and $B^*K$ thresholds.
As a consequence, the only allowed strong decay to
$B_s^* \pi$ would not preserve isospin and the resonances 
are expected to
be narrow, of the order of a few keV. A second 
$b\bar s$ $J=1^+$ resonance appears at $M=5857$ MeV,
with almost 99\% of $q\bar q$ component which may correspond 
with the new $B_{s1}$ state reported by CDF with a mass around 5829 MeV.
The fourth state appears at 
6174 MeV. A similar calculation in the non-strange sector 
is much more involved and we can only predict the existence 
of a $B^*_{0}$ state with $M=5615$ MeV and 48\% of four-quark component 
and 51\% of $b\bar n$ pair. The lowest state, representing
the $B^*_0$, is 200 MeV above the isospin preserving threshold $B\pi$, 
therefore it would be broad. The orthogonal state appears higher in energy, 
at 6086 MeV, also with an important four-quark component.

Our results are in agreement with Refs.~\cite{Bardeen} 
and~\cite{col} obtained with different approaches. Ref.~\cite{Bardeen} interprets the two 
$D_{sJ}$ resonances as the chiral partners of the $(0^-,1^-)$ ground state spin multiplet.
Fitting the chiral mass gap between the $(0^-,1^-)$ and $(0^+,1^+)$ multiplets to the 
$D_{sJ}^*(2317)$ experimental mass, they reproduce the $D_{sJ}^*(2460)$ mass. 
Assuming that in the $M_Q\rightarrow\infty$ limit the chiral mass gap for the 
charm and bottom sectors is the same, they predict the 
masses for the $B$ meson chiral multiplet $(0^+,1^+)$. 
In Ref.~\cite{col} the authors impose invariance under heavy quark spin-flavor and chiral transformations
to build an effective QCD lagrangian. The effective parameters are determined using
as inputs the experimental $D$, $D_s$ and $B_s$ masses together with the assumption that the
mass splitting between positive and negative parity 
doublets is the same in the charm and bottom sectors.
Their results are summarized in Table~\ref{t2}.
These calculations strengthen the idea that the anomalies 
observed in the charm sector must appear in the bottom one.

Our approach is similar to Ref.~\cite{lee} considering one loop chiral corrections to calculate 
the $D_{sJ}^*(2317)$ mass. Such corrections lower the mass of the $0^+$ state and therefore 
can account for the unusually small mass of the $D_{sJ}^*(2317)$ and the small
mass difference between $D_{sJ}^*(2317)$ and $D^*_0(2308)$. However, our approach
has the additional property that considering 
four-quark configurations, besides playing a similar role to the one loop corrections, 
it augments the number of states. This provides a plausible interpretation
of the recently measured $D_{s}^*(2860)$ \cite{au} as the orthogonal partner 
of the mixture of conventional $P$ wave 
quark-antiquark states and four-quark components describing the $D_{sJ}^*(2317)$ \cite{FB}.

In the L3 Collaboration results~\cite{L3} the masses of the
$B^*_0$ and $B_1$ mesons, although not explicitly given, are constrained by 
heavy quark effective theory relations~\cite{Man00}, 
$M(B_2^*)-M(B_1)\approx M(B_1')-M(B_0^*)\approx 12$ MeV. This allows to 
estimate the masses of the $B^*_0$ and $B_1$ mesons from the 
$B^\prime_1$ and $B^*_2$ experimental masses, 
obtaining $M(B^*_0)\approx5658$ MeV and $M(B_1)\approx5756$ MeV.
The $B_1$ mass value agrees with the recent measurement of 
CDF and D0 Collaborations~\cite{Pau} and with our prediction (see 
Table~\ref{t3}) whereas the $B^*_0$ mass is very close to our $I=1/2$,
$J^P=0^+$ state with a 48\% of four-quark component (see Table~\ref{t4}).
Furthermore the L3 Collaboration observed an excess of events above the 
expected background in the 5.9-6.0 GeV region of the $B\pi$ mass spectrum,
what might correspond to our second $I=1/2$, $J^P=0^+$ mixed state at 6086 MeV.
The L3 data seem to indicate that 
the $L=1$ excited $B$ mesons show the same behavior as
the corresponding excited states in the open charm sector.

The isovector four-quark states do not couple to 
the $b\bar s$ system and they are therefore much 
higher in energy, being both close to 6 GeV and 
above the strong decay threshold. Although they 
should be very broad, and therefore very difficult 
to observe, any indication of its existence 
would be a definite signal of the presence 
of four-quark states in the heavy-meson sector.

The structure of the excited $B$ mesons can also be
analyzed studying their 
electromagnetic decay widths. The formalism needed 
to evaluate the $\Gamma[b\bar q\to b \bar q+\gamma$], $\Gamma[bn\bar q
\bar n\to bn \bar q\bar n+\gamma$], and $\Gamma[bn\bar q\bar n\to b 
\bar q+\gamma]$ widths has been described in 
Ref. \cite{Vij73}. We compare in Table~\ref{t5} our results for the 
radiative transitions of the $0^+$ and $1^+$ states for the 
cases of a pure $b\bar q$ structure, QM($b\bar q$), a mixed one, 
QM($b\bar q+bq\bar q\bar q$), and those of Ref.~\cite{Bardeen}. 
The difference between the QM($b\bar q$) results and those
of Ref.~\cite{Bardeen} are due to phase space.
The difference of the previous two cases with the 
QM($b\bar q+bq\bar q\bar q$) results
can be traced back to the more involved wave function 
of this mixed case. The mixing among the $^3P(b\bar q)$,
$^1P(b\bar q)$ and the four-quark
components generates a very small $^3P(b\bar q)$
probability for some particular states, as for example
the $J^P=1^+$, of the order of 1\%. The electromagnetic
decay of the $^1P(b\bar q)$ component to the $J^P=1^-$ 
state is forbidden.
The transition from the four-quark component to the
meson-photon state does only occur for a very
particular component of the tetraquark wavefunction: the
one where the light quark-antiquark pair is in a
color singlet, spin one, isospin zero state, with a
very small probability for the state under consideration.
The most significant consequence is the suppression 
predicted for the $1^+ \to 1^- + \gamma$ decay as compared to the 
$1^+ \to 0^- + \gamma$. 
A ratio ${{1^+ \to 0^- + \gamma}\over{1^+ \to 1^- + \gamma}}\approx 1 $	
has been obtained in Ref.~\cite{Bardeen} 
and in our results with a pure $b\bar q$ structure. For the mixed case 
we find a much larger value for this ratio, 
${{1^+ \to 0^- + \gamma}\over{1^+ \to 1^- + \gamma}}\approx 100 $	
due to the small 
$1^3P_1$ $c\overline s$ probability of the $1^+$ state. In view of 
these predictions, once experimentally measured, the electromagnetic 
decay widths would be an important diagnostic tool 
to understand the nature of these states.

In brief, we have performed an exploratory study of 
the $L=1$ excited $B$ mesons in terms of two- and 
four-quark components based in our experience 
on the open-charm mesons.
Our results agree with the recently measured $B$ meson states
by CDF and D0 Collaborations. In addition we predict 
the existence of two resonances, $B^*_{s0}$ and $B^\prime_{s1}$,
with almost 30\% of four-quark component, 
which lie below the $BK$ and $B^*K$ thresholds, respectively. 
Thus, the only allowed strong decays would violate isospin 
and the resonances would be narrow. In the non-strange sector we 
did not find such narrow resonances but our results give support 
to the L3 Collaboration findings. Therefore the 
mixing between two and four-quark components, which 
explains the unexpected low masses of $D_{sJ}^*(2317)$ and 
$D_{sJ}(2460)$ open-charm states, would also play a relevant role 
in the open-beauty sector. We also found that the
ratio ${{1^+ \to 0^- + \gamma}\over{1^+ \to 1^- + \gamma}}$	
would provide an experimental signature of the proposed structure.  

We encourage experimentalists on the measurement of 
the spectroscopic and electromagnetic properties 
of the positive parity $B_J$ and $B_{sJ}$ states, 
that would help to clarify the exciting situation 
of the open-bottom and open-charm mesons.

\vspace*{1cm}

This work has been partially funded by Ministerio de Ciencia y Tecnolog\'{\i}a
under Contract No. FPA2007-65748 and by Junta de Castilla y Le\'{o}n
under Contract No. SA016A07.

\begin{table}
\caption{Expected decay properties of the $P$ wave $B_s$ mesons. 
{\it J} is the total angular momentum, $P$
its parity, and {\it $j_q$} the total angular momentum of the light quark.}
\label{t1}
\begin{center}
\begin{tabular}{|cc|ccc|}
\hline
&$j_q$ $J^P$&State &Decay mode	&Width \\
\hline
&$1/2$ $0^+$&$B^*_{s0}$&$BK$  &broad\\
&$1/2$ $1^+$&$B^{'}_{s1}$& $B^*K$  &broad\\
&$3/2$ $1^+$&$B_{s1}$& $B^*K$  &narrow\\
&$3/2$ $2^+$&$B^*_{s2}$	& $BK$, $B^*K$  &narrow \\
\hline
\end{tabular}
\end{center}
\end{table}

\begin{table}
\caption{$b\overline n$ and $b\overline s$ spectra, in MeV, using different approaches.}
\label{t2}
\begin{center}
\begin{tabular}{|c|ccccccc|}
\hline
$b\overline n$&Ref.~\cite{Ebert}&Ref.~\cite{God}&Ref.~\cite{Bardeen}
&Ref.~\cite{col}&Ref.~\cite{Falk}&Ref.~\cite{Eich}&Ref.~\cite{green}\\
\hline
$B^*_0$& 5738		& 5760		& 5627&5700	&$-$	&$-$&$-$\\
$B^{'}_1$& 5719		& 5780		& 5674&5750&$-$	&$-$&$-$\\
$B_1$& 5757		& 5780		& $-$& 5774&5780&5755&$-$\\
$B^*_2$& 5733		& 5800		& $-$& 5790&5846&5767 &$-$\\
\hline
\hline
$b\overline s$&Ref.~\cite{Ebert}&Ref.~\cite{God}&Ref.~\cite{Bardeen}
&Ref.~\cite{col}&Ref.~\cite{Falk}&Ref.~\cite{Eich}&Ref.~\cite{green}\\
\hline
$B^*_{s0}$& 5841		& 5830		& 5718&5710&$-$&$-$&5756$\pm$31\\
$B^{'}_{s1}$& 5831		& 5786		& 5765&5770&$-$&$-$&5804$\pm$31\\
$B_{s1}$& 5859		& 5786		& $-$& 5877	&5886&5834&5892$\pm$52\\
$B^*_{s2}$& 5844		& 5808		& $-$& 5893		&5899&5846&5904$\pm$52\\
\hline
\end{tabular}
\end{center}
\end{table}

\begin{table}
\caption{$b\overline s$ and $b\overline n$ quark model (QM) masses, in MeV.
Experimental data (Exp.) are from
Ref. \protect\cite{PDG}, except for the states denoted by a dagger
from Ref. \protect\cite{Pau} and by a double dagger
from Ref.~\protect\cite{L3}.}
\label{t3}
\begin{center}
\begin{tabular}{|cc|cc||cc|}
\hline
&$nL$ $J^P$	&QM $(b\bar s)$ 	&Exp.		&QM $(b\bar n)$	&Exp.\\
\hline
&$1S$ $0^-$	&5355  			&5369.6$\pm$2.4	&5281 		&5279.2$\pm$0.5\\
&$1S$ $1^-$	&5400  			&5416.6$\pm$3.5	&5321 		&5325.0$\pm$0.6\\
&$1P$ $0^+$	&5838  			&$-$ 		&5848 		&$-$\\
&$1P$ $1^+$	& 
$\left.\begin{array}{c} 5837 \\ 5869 \end{array}\right\}$ 
& 5829.4$\pm0.2\pm0.6^{\dagger}$ &
$\left.\begin{array}{c} 5768 \\5876\end{array}\right\}$ & 
$\left\{\begin{array}{c} 5734\pm 3\pm2^{\dagger}\\5720.8\pm 2.5\pm5.3^{\dagger}\\
5670\pm 10\pm13^{\dagger\dagger}\end{array}\right.$\\
&$1P$ $2^+$	&5853  			&
$\left\{\begin{array}{c} 5839.6\pm 0.4 \pm0,5^{\dagger} \\ 5839.1\pm 1.4\pm1.5^{\dagger}\end{array}\right.$	
& 5786	&$\left\{\begin{array}{c}
5738\pm 6\pm1^{\dagger}\\5746.0\pm 2.5\pm5.3^{\dagger}\\5768\pm 5\pm6^{\dagger\dagger}\end{array}\right.$\\
\hline
\end{tabular}
\end{center}
\end{table}

\begin{table}
\caption{Masses (QM), in MeV, and probability of
the different wave function components for $B_s$ mesons 
once the mixing between $b\bar q$ and $bq\bar q\bar q$ configurations 
is considered.}
\label{t4}
\begin{center}
\begin{tabular}{|c|cc||c|cc||c|cc|}
\hline
\multicolumn{6}{|c||}{$I=0$} & \multicolumn{3}{|c|}{$I=1/2$} \\
\hline
\multicolumn{3}{|c||}{$J^P=0^+$}    & \multicolumn{3}{|c||}{$J^P=1^+$} & \multicolumn{3}{|c|}{$J^P=0^+$} \\
\hline
QM                  &5679   	&6174	&QM			&5713		&5857  		
&QM                   &5615 	&6086 \\
\hline
P($bn\bar s\bar n$) &0.30	&0.51  	&P($bn\bar s\bar n$)	&0.24  		&$\sim 0.01$ 	
&P($bn\bar n\bar n$)  &0.48       &0.46  \\
P($b\bar s_{1^3P}$) &0.69   	&0.26  	&P($b\bar s_{1^1P}$)	&0.74  		&$\sim 0.01$ 	
&P($b\bar n_{1P}$)    &0.51       &0.47 \\
P($b\bar s_{2^3P}$) &$\sim 0.01$	&0.23  	&P($b\bar s_{1^3P}$)	&$\sim 0.01$ 	&0.99		
&P($b\bar n_{2P}$)    &$\sim 0.01$ &0.07 \\
\hline
\end{tabular}
\end{center}
\end{table}

\begin{table}
\caption{Comparison of radiative decay widths (keV) 
assuming only a pure $b\bar q$ structure, QM($b\bar q$), 
a combination of two- and four-quark components, QM($b\bar q+bq\bar q\bar q$), 
and those of Ref. \cite{Bardeen}.}
\label{t5}
\begin{center}
\begin{tabular}{|c|ccc|}
\hline
Transition	& QM($b\bar q$)	& QM($b\bar q+bq\bar q\bar q$)	&Ref.~\cite{Bardeen}\\
\hline
$0^+ \to 1^- + \gamma$	& 171.4		& 31.9				& 58.3	\\
$1^+ \to 1^- + \gamma$	& 75.6		& 0.6 				& 56.9  \\
$1^+ \to 0^- + \gamma$	& 106.5		& 60.7				& 39.1	\\
\hline
${{1^+ \to 0^- + \gamma}\over{1^+ \to 1^- + \gamma}}$	
				& 1.41		& 101.17	& 0.69	\\
\hline
\end{tabular}
\end{center}
\end{table}
\end{document}